\documentclass[epj]{svjour}
\usepackage{latexsym}
\usepackage{graphics}
\usepackage[utf8]{inputenc}
\usepackage{verbatim}
\usepackage{comment}
\usepackage[T1]{fontenc}
\usepackage{epsfig}
\usepackage{bm}
\usepackage{xcolor}
\newcommand\rf[1]{(\ref{eq:#1})}
\newcommand\lab[1]{\label{eq:#1}}
\newcommand\nonu{\nonumber}
\newcommand\br{\begin{eqnarray}}
\newcommand\er{\end{eqnarray}}
\newcommand\be{\begin{equation}}
\newcommand\ee{\end{equation}}

\newcommand\lb{\lbrack}
\newcommand\rb{\rbrack}

\newcommand\bc{\begin{center}}
\newcommand\ec{\end{center}}

\renewcommand\a{\alpha}

\renewcommand\d{\delta}
\newcommand\eps{\epsilon}
\newcommand\vareps{\varepsilon}

\newcommand\G{\Gamma}

\newcommand\h{\frac{1}{2}}
\renewcommand\k{\kappa}
\renewcommand\l{\lambda}
\renewcommand\L{\Lambda}
\newcommand\m{\mu}
\newcommand\n{\nu}
\newcommand\om{\omega}

\newcommand\vp{\varphi}
\renewcommand\P{\Phi}
\newcommand\pa{\partial}

\newcommand\pr{\prime}

\newcommand\s{\sigma}

\newcommand\wti{\widetilde}

\newcommand\cB{{\mathcal B}}

\newcommand\cL{{\mathcal L}}
\newcommand\cM{{\mathcal M}}
\newcommand\cN{{\mathcal N}}

\newcommand\cT{{\mathcal T}}

\newcommand{\ct}[1]{\cite{#1}}
\begin{document}
\title{Non-Canonical Volume-Form Formulation of Modified Gravity Theories and Cosmology}
\author{David Bensity \inst{1} \and Eduardo I. Guendelman \inst{1,2,3} \and Alexander Kaganovich \inst{1} \and Emil Nissimov\inst{4} \and Svetlana Pacheva\inst{4}
\thanks{\emph{Present address:} Insert the address here if needed}%
}                     
%
%
\institute{Physics Department, Ben-Gurion University of the Negev, Beer-Sheva 84105, Israel \and Frankfurt Institute for Advanced Studies (FIAS), Ruth-Moufang-Strasse~1, 60438 Frankfurt am Main, Germany \and Bahamas Advanced Study Institute and Conferences, 4A Ocean Heights, Hill View Circle, Stella Maris, Long Island, The Bahamas \and Institute for Nuclear Research and Nuclear Energy, Bulgarian Academy
of Sciences, Sofia, Bulgaria}
%
%
\abstract{
A concise description is presented of the basic features of the formalism of non-canonical 
spacetime volume-forms and its application in modified gravity theories and
cosmology. The well known unimodular gravity theory appears as a very special case. 
Concerning the hot issues facing cosmology now, we specifically briefly outline 
the construction of:
(a) unified description of dark energy and dark matter as manifestations of 
a single material entity -- a special scalar field ``darkon''; 
(b) quintessential models of universe evolution with a gravity-``inflaton''-assisted 
dynamical Higgs mechanism -- dynamical suppression/generation of spontaneous 
electroweak gauge symmetry breaking in the ``early''/``late'' universe; 
(c) unification of dark energy and dark matter with diffusive interaction among them; 
(d) mechanism for suppression of 5-th force without fine-tuning.
\PACS{
      {PACS-key}{discribing text of that key}   \and
      {PACS-key}{discribing text of that key}
     } 
} 
\maketitle
\section{Non-Riemannian Volume-Form Formalism}
\label{intro}

A broad class of actively developed modified/extended gravitational theories is based on
employing alternative non-Riemannian spacetime volume-forms (metric-independent 
generally covariant volume elements) in the pertinent Lagrangian actions instead 
of, or alongside with, the canonical Riemannian volume element
given by the square-root of the determinant of the Riemannian metric. This
method was originally proposed in \ct{Guendelman:1999qt,Guendelman:1999tb} and for a concise 
geometric formulation using differential forms combined 
with canonical Hamiltonian formalism for systems with constraints (gauge symmetries), 
see \ct{Guendelman:2014lea,Guendelman:2015qva} (an earlier geometric formulation with a ``quartet'' 
of scalar fields appeared in \ct{Gronwald:1997ei}).

Volume-forms are fairly basic objects in differential geometry -- they exist
on arbitrary differentiable manifolds and define covariant (under general
coordinate reparametrizations) integration measures. It is important to
stress that the existence of volume-forms is {\em completely independent} of the
presence or absence of additional geometric structures on the manifold  -- 
Volume forms are defined \ct{spivak} by nonsingular maximal rank differential forms $\om$:
\br
\int_{\cM} \om \bigl(\ldots\bigr) = \int_{\cM} dx^D\, \Omega \bigl(\ldots\bigr)
\;\; ,
\nonu \\
\om = \frac{1}{D!}\om_{\m_1 \ldots \m_D} dx^{\m_1}\wedge \ldots \wedge dx^{\m_D}\; ,
\lab{omega-1} \\
\om_{\m_1 \ldots \m_D} = - \vareps_{\m_1 \ldots \m_D} \Omega \; ,
\nonu
\er
(our conventions for the alternating symbols $\vareps^{\m_1,\ldots,\m_D}$ and
$\vareps_{\m_1,\ldots,\m_D}$ are: $\vareps^{01\ldots D-1}=1$ and
$\vareps_{01\ldots D-1}=-1$).
The volume element density 
$\Omega$ transforms as scalar density under general coordinate reparametrizations.

In standard generally-covariant theories (with action $S=\int d^D\! x \sqrt{-g} \cL$)
the Riemannian spacetime volume-form is defined through the ``D-bein''
(frame-bundle) canonical one-forms $e^A = e^A_\m dx^\m$ ($A=0,\ldots ,D-1$):
\br
\om = e^0 \wedge \ldots \wedge e^{D-1} = \det\Vert e^A_\m \Vert\,
dx^{\m_1}\wedge \ldots \wedge dx^{\m_D} 
\longrightarrow \quad
\Omega = \det\Vert e^A_\m \Vert\, d^D x = \sqrt{-\det\Vert g_{\m\n}\Vert}\, d^D x \; .
\lab{omega-riemannian}
\er

Instead of, or alongside with,  $\sqrt{-g}$ we can employ one or several different  
alternative {\em non-Riemannian} volume elements as in \rf{omega-1} given by non-singular 
{\em exact} $D$-forms $\om^{(j)} = d B^{(j)}$ where:
\br
B^{(j)} = \frac{1}{(D-1)!} B^{(j)}_{\m_1\ldots\m_{D-1}}
dx^{\m_1}\wedge\ldots\wedge dx^{\m_{-1}} 
\longrightarrow \quad  \Omega^{(j)} \equiv \Phi(B^{(j)}) =
\frac{1}{(D-1)!}\vareps^{\m_1\ldots\m_D}\, \pa_{\m_1} B^{(j)}_{\m_2\ldots\m_D} \; .
\lab{Phi-D}
\er
In other words, the non-Riemannian volume elements are defined in terms of
the dual field-strengths of auxiliary rank $D-1$ tensor gauge fields 
$B^{(j)}_{\m_1\ldots\m_{D-1}}$. 

Let us again strongly emphasize that the term
``non-Riemannian'' concerns only the nature of the non-canonical volume
elements, which exist on the spacetime manifold with a standard Riemannian
geometric structure, 
torsionless affine connection $\G^\l_{\m\n}$ either independent of $g_{\m\n}$
(first-order metric-affine / Einstein-Palatini formalism) or as a Levi-Civita connection 
w.r.t. $g_{\m\n}$ (second-order purely metric / Einstein-Hilbert formalism).

The generic form of modified gravity actions involving (one or more) non-Riemannian 
volume-elements, called for short 
actions, read (henceforth $D=4$, and we will use units with 
$16\pi G_{Newton} =1$):
\br
S = \int d^4 x \,\P(B^{(1)}) \bigl( R +\cL^{(1)}\bigr) 
\nonu + \int d^4 x \,\sum_{j\geq 2} \P(B^{(j)})\, \cL^{(j)} + \int d^4 x \,\sqrt{-g}\cL^{(0)} \; ,
\lab{NRVF-0}
\er
where $R$ is the scalar curvature. 
The equations of motion of \rf{NRVF-0} w.r.t. the auxiliary tensor gauge fields 
$B^{(j)}_{\m\n\k}$ according to \rf{Phi-D} imply:
\br
\pa_\m \bigl( R+\cL^{(1)}\bigr)=0 \;\; ,\;\; 
\pa_\m \cL^{(j)} = 0 \;\; (j\geq 2) \; ,\, \longrightarrow \; \; R+\cL^{(1)}=M_1 \quad,\quad \cL^{(j)}= M_j  \; ,
\lab{L-M}
\er
where all $M_j$ ($j\geq 1$) are {\em free integration constants} not present in the 
original NRVF gravity action \rf{NRVF-0}. 

A characteristic feature of the 
NRVF gravitational theories \rf{NRVF-0} is that
when starting in the first-order (Palatini) formalism all non-Riemannian
volume-elements $\P(B^{(j)})$ yield almost {\em pure-gauge} degrees of freedom, 
additional physical (field-propagating) gravitational degrees of freedom 
except for few discrete degrees 
of freedom with conserved canonical momenta appearing as the arbitrary integration 
constants $M_j$ in \rf{L-M}. 
The reason is that the NRVF gravity action \rf{NRVF-0} in Palatini
formalism is linear w.r.t. the velocities of some of the components of the 
auxiliary gauge fields $B^{(j)}_{\m\n\k}$  defining the non-Riemannian 
volume-element densities, and does not depend on the velocities of the rest of 
auxiliary gauge field components. 
The (almost) pure-gauge nature of the latter is explicitly shown 
in \ct{Guendelman:2015qva,Guendelman:2016lea} (appendices A) employing the standard 
canonical Hamiltonian treatment of systems with gauge symmetries, i.e.,
systems with first-class Hamiltonian constraints a'la Dirac 
\ct{henneaux-teitelboim,rothe}.

However, in the second-order formalism 
(where $\G_{\m\n}^\l$ is the usual Levi-Civita connection w.r.t. $g_{\m\n}$)
the first non-Riemannian volume form $\P(B^{(1)})$ in \rf{NRVF-0}
is {\em not} any more pure-gauge. The reason is that the scalar curvature $R$ 
(in the metric formalism) contains {\em second-order} 
(time) derivatives (the latter amount to a total derivative in the ordinary case 
$S= \int d^4 x \sqrt{-g} R + \ldots$). Now defining $\chi_1 \equiv \P(B^{(1)})/\sqrt{-g}$, 
the latter field becomes physical degree of freedom as seen from the equations of motion
of \rf{NRVF-0} w.r.t. $g^{\m\n}$:
\be
R_{\m\n} + \frac{1}{\chi_1} \bigl(g_{\m\n}\Box\chi_1 
-\nabla_\m \nabla_\n \chi_1 \bigr) + \ldots = 0 \; .
\lab{chi1-degree-freedom}
\ee

As a final introductory remark let us note that the well-known covariant
formulation of unimodular gravity \ct{Henneaux:1989zc} can be viewed as a
simple particular case within the general class \rf{NRVF-0} of modified gravity
actions based on the non-Riemannian volume-form formalism. Indeed, the original action of unimodular gravity \cite{Henneaux:1989zc} reads:
\begin{equation}
\mathcal{S} = \int d^4 x \sqrt{-g} \left(R + 2 \Lambda + \mathcal{L}_m\right) 
- \int d^4x \, \Phi\, 2\Lambda 
\end{equation}
with $\Lambda$ being a dyamical field, and $\Phi \equiv \pa_\m F^\m$ where
the vector density $F^\m$ can be written as Hodge-dual
$F^\m \equiv \frac{1}{3!}\vareps^{\m\n\k\l} B_{\n\k\l}$
w.r.t. rank 3 auxiliary gauge field $B_{\n\k\l}$
(cf. \rf{Phi-D} for $D=4$). Variation w.r.t. $F^\m$ implies $\Lambda = const$,
whereas variation w.r.t. $\Lambda$ yields $\Phi = \sqrt{-g}$, 
in what follows, for general NRVF gravity models \rf{NRVF-0} the field ratio
$\chi_1$ is either a non-trivial algebraic function of the matter fields in 
$\cL^{(j)}$ within the first-order (Palatini) formalism (cf. Eq.\rf{chi-1} below), 
or it becomes a new dynamical scalar field within the second-order (metric) formalism 
(cf. Eq.\rf{chi1-degree-freedom}). 
\section{Simple Model of Unified Dark Energy and Dark matter}
\label{sec:1}
A simple NRVF gravity model providing a unified description of dark energy and 
dark matter defined by an action, particular representative of the class \rf{NRVF-0}, 
was proposed in \ct{Guendelman:2014bva,Guendelman:2015jii}:
\be
S = \int d^4 x\, \Bigl\lb\sqrt{-g} \bigl(R + X-V_1(\phi)\bigr) 
+  \Phi(B) \bigl( X-V_2(\phi)\bigr)\Bigr\rb \; ,
\lab{NRVF-1a}
\ee
or equivalently:
\br
S = \int d^4 x\, \sqrt{-g} \bigl(R - U(\phi)\bigr) 
+ \bigl(\sqrt{-g}+\Phi(B) \bigr)\bigl(X - V(\phi)\bigr)
\nonu \\
\phantom{}\lab{NRVF-1}
\er
using the notations: 
$V\equiv V_2,\;\; U\equiv V_1 - V_2,\, X\equiv -\h g^{\m\n}  \pa_\m \phi \pa_\n\phi$, 
and $\Phi(B) \equiv 1/3! \vareps^{\m\n\k\l} \pa_\m B_{\n\k\l}$ (cf. \rf{Phi-D}). 
Variation of the action \rf{NRVF-1} w.r.t. auxiliary gauge field $B_{\n\k\l}$
yields (cf. the general Eqs.\rf{L-M}):
\be
X - V(\phi) = - 2M_0, 
\lab{L-const}
\ee
where $M_0$ is free integration constant. The variation of \rf{NRVF-1} w.r.t. 
scalar field $\phi$ can be written in the following suggestive form:
\br
\nabla_\m J^\m = - \sqrt{2X} U^\pr (\phi) \; ,
\lab{J-nonconserv} \\
J_\m \equiv - \bigl(1+\chi\bigr)\sqrt{2X} \pa_\m \phi \;\; ,\;\;
\chi \equiv \frac{\Phi(B)}{\sqrt{-g}} \; .
\lab{J-current} 
\er
The dynamics of $\phi$ is entirely determined by the dynamical constraint \rf{L-const}, 
completely independent of the potential $U(\phi)$. 
On the other hand, the $\phi$-equation of motion written in the form 
\rf{J-nonconserv} is in fact an equation determining the dynamics of $\chi$. 
The energy-momentum tensor $T_{\m\n}$ in the Einstein equations can be
written in a relativistic hydrodynamical form as:
\be
T_{\m\n} = \rho_0 u_\m u_\n + g_{\m\n} {\wti p}\;\;,\quad J_\m = \rho_0 u_\m
\lab{EM-hydro}
\ee
where $u_\m$ is a fluid velocity unit vector:
\be
u_\m \equiv - \frac{\pa_\m \phi}{\sqrt{2X}} \quad 
({\rm note} \; u^\m u_\m = -1\;) \; ,
\lab{fluid-velocity}
\ee
and the energy density ${\wti \rho}$ and pressure ${\wti p}$ are given as:
\be
{\wti \rho} = \rho_0 + 2M_0 + U(\phi),\quad {\wti p} = -2M_0 - U(\phi)
\lab{fluid-rho-p}
\ee
with $\rho_0 \equiv (1+\chi) 2X = {\wti \rho} + {\wti p}$. 
Energy-momentum conservation $\nabla^\n T_{\m\n}= 0$ implies:
\be
\nabla^\m \bigl(\rho_0 u_\m\bigr) = - \sqrt{2X}\,U^\pr(\phi), \quad
u_{\n} \nabla^{\n} u_\m =0 \; , 
\lab{EM-conserve}
\ee
the last Eq.\rf{EM-conserve} meaning that the matter fluid flows along geodesics. In Eqs.\rf{EM-hydro}, \rf{fluid-rho-p} the quantity $\rho_{\rm DE} \equiv 2M_0 + U(\phi) = - {\wti p}$ has the interpretation as dark energy density, whereas $\rho_0$ is the dark matter energy density. For $U(\phi) = {\rm const}$ or $U(\phi)=0$ the model \rf{NRVF-1} possesses a non-trivial hidden nonlinear Noether symmetry under:
\br
\d_\eps \phi = \eps \sqrt{X},\quad \d_\eps g_{\m\n} = 0 \; , \quad
\d_\eps \cB^\m = - \eps \frac{1}{2\sqrt{X}} \phi^{,\mu} 
\bigl(\P(B) + \sqrt{-g}\bigr)  \; ,
\lab{hidden-sym}
\er
where $\cB^\m \equiv \frac{1}{3!} \vareps^{\m\n\k\l} B_{\n\k\l}$, with a 
Noether conserved current $J^\m = \rho_0 u_\m$ according to \rf{J-current}:
$\nabla_\m J^\m = 0$. Specifically, for Friedmann-Lemaitre-Robertson-Walker 
metric with Friedmann scale factor $a(t)$ Eq.\rf{J-current} with $U(\phi)=0$
implies: $\rho_0 = c_0 /a^3$, $c_0$ being a free integration constant.

Thus, according to \rf{EM-hydro}, \rf{fluid-rho-p}
the model  provides an exact description of $\Lambda$CDM model, and for a
non-trivial potential $U(\phi)$, breaking the hidden Noether symmetry \rf{hidden-sym},
we have interacting dark energy and dark matter.

The above interpretation justifies the alias ``darkon'' for the scalar field $\phi$.
Let us specifically emphasize that both dark energy and dark matter components of 
the energy density \rf{fluid-rho-p} have been {\em dynamically} generated thanks 
to the non-Riemannian volume element construction -- both due to the appearance of the 
free integration constant $M_0$ and of the hidden nonlinear Noether symmetry
\rf{hidden-sym} (``darkon'' symmetry). 
In Ref.\ct{Benisty:2020nql} the correspondence between $\Lambda$CDM model and the
``darkon'' Noether symmetry was exhibited up to linear order w.r.t.
gravity-matter perturbations and the implications of the ``darkon'' symmetry
breaking for possible explanation of the cosmic tensions was briefly discussed. 
 \cite{Staicova:2016pfd} confront some potential with the late accelerated expansion data. 
\section{Quintessential Inflationary Model with Dynamical Higgs Effect in Metric-Affine Formulation}

The starting point is the following specific NRVF gravity action from the
class \rf{NRVF-0} involving coupling to a scalar ``inflaton'' $\vp$ and to the
bosonic sector of the standard electroweak particle model where, following
the remarkable Bekenstein's idea from 1986 \ct{Bekenstein:1986hi} about
gravity-assisted dynamical spontaneous symmetry breakdown, the Higgs-like 
$SU(2)\times U(1)$ iso-doublet scalar $\s_a$ enters with a standard positive
mass-squared and without self-interaction in sharp distinction w.r.t.
standard particle model. The pertinent NRVF action reads explicitly 
\ct{Guendelman:2016lea,Brink:2020fie,Benisty:2020vvm}:
\br
S = \int d^4 x\,\P_1 (A)\Bigl\lb R(g,\G) - 2 \L_0 \frac{\P_1 (A)}{\sqrt{-g}} 
+ L^{(1)}(\vp,\s) \Bigr\rb + \int d^4 x\,\P_2 (B)\Bigl\lb f_2 e^{2\a\vp} + L_{\rm EW-gauge}
- \frac{\P_0 (C)}{\sqrt{-g}}\Bigr\rb
\; ,
\lab{NRVF-2}
\er
with notations:
\begin{itemize}
\item 
$\P_1 (A) = \frac{1}{3!}\vareps^{\m\n\k\l}\, \pa_{\m} A_{\n\k\l}$ and
similarly for $\P_2 (B)$, $\P_0 (C)$ according to \rf{Phi-D};
\item 
The scalar curvature $R(g,\G) = g^{\m\n} R_{\m\n}(\G)$ is given in terms
of the Ricci tensor $R_{\m\n}(\G)$ in the first-order (Palatini) formalism;
\item
The matter Lagrangian reads:
\br
L^{(1)} (\vp,\s) = X_\vp + f_1 e^{\a\vp} + X_\s - m_0^2 \s^{*}_a\s_a e^{\a\vp},\quad
X_\vp \equiv - \h g^{\m\n} \pa_\m \vp \pa_\n \vp \;\;,\;\;
X_\s \equiv - g^{\m\n} \nabla_\m \s^{*}_a \nabla_\n \s_a \; ;
\lab{L-1} 
\nonu
\er
\item
$L_{\rm EW-gauge}$ denotes the Lagrangian of the $SU(2)\times U(1)$ gauge
fields.
\item
$\L_0$ is small dimensionful constant which will be identified in
the sequel with the ``late'' universe cosmological constant in the dark
energy dominated accelerated expansion's epoch.
\end{itemize}

The equations of motion w.r.t. auxiliary tensor gauge fiels in $\P_1 (A)$, $\P_2 (B)$ and 
$\P_1 (C)$ yield (cf. \rf{L-M}):
\br
g^{\m\n}\Bigl(R_{\m\n}(\G) - \h \pa_\m \vp \pa_\n \vp
- \nabla_\m \s^{*}_a \nabla_\n \s_a \Bigr) - 4\L_0 \frac{\P_1(A)}{\sqrt{-g}}
+ \bigl(f_1 - m^2_0\, \s^{*}_a \s_a\bigr) e^{\a\vp} = M_1, 
\lab{M1} \\
f_2 e^{-2\a\phi} + L_{\rm EW-gauge} - \frac{\P_0 (C)}{\sqrt{-g}} = - M_2, \,\, \frac{\P_2 (B)}{\sqrt{-g}} = \chi_2 
\lab{M2} 
\lab{chi2}
\er
where $M_{1,2}, \chi_2$ are integration constants. The $g^{\m\n}$-equations of motion together with \rf{M1}-\rf{chi2} imply that the ratio $\chi_1 \equiv \frac{\P(A)}{\sqrt{-g}}$ is an algebraic
function of the matter fields:
\be
\chi_1 (\vp,\s) \equiv \frac{\P_1(A)}{\sqrt{-g}}= 
\frac{2\chi_2\bigl(f_2 e^{2\a\vp}+M_2\bigr)}{M_1 +
\bigl(m^2_0\, \s^{*}_a \s_a - f_1\bigr) e^{\a\vp}}.
\lab{chi-1}
\ee

The equation of motion w.r.t. 
$\G^\l_{\m\n}$, following analogous derivation in \ct{Guendelman:1999qt},  
yields a solution for $\G^\m_{\n\l}$ as a Levi-Civita connection
w.r.t. to a {\em Weyl-conformally rescaled} metric: 
\be
{\bar g}_{\m\n} = \chi_1 (\vp,\s)\, g_{\m\n} 
\lab{g-bar}
\ee
with $\chi_1 (\vp,\s)$ as in \rf{chi-1}. The conformal transformation $g_{\m\n} \to {\bar g}_{\m\n}$ via \rf{g-bar} on the NRVF action \rf{NRVF-2} converts the latter into the physical Einstein-frame action (objects in the Einstein-frame are indicated by a bar):
\br
S_{\rm EF} = \int d^4 x\,\sqrt{-{\bar g}} \Bigl\lb R({\bar g})
- \h {\bar g}^{\m\n} \pa_\m \phi \pa_\n \phi - {\bar g}^{\m\n} \nabla_\m \s^{*}_a \nabla_\n \s_a 
- U_{\rm eff}(\vp,\s) + L_{\rm EW-gauge}({\bar g})\Bigr\rb \; .
\lab{NRVF-2-EF}
\er
Here the interesting object is the effective Einstein-frame scalar potential:
\be
U_{\rm eff}(\vp,\s)  = \frac{\Bigl\lb M_1 + e^{\a\vp}\bigl(m^2_0\, \s^{*}_a \s_a 
- f_1\bigr)\Bigr\rb^2}{4\chi_2 \bigl(f_2 e^{2\a\vp}+M_2\bigr)} + 2\L_0 \; ,
\lab{U-eff}
\ee
which is entirely {\em dynamically generated} due to the appearance of the
free integration constants $M_{1,2}$ and $\chi_2$ \rf{M1}-\rf{chi2}. $U_{\rm eff}(\vp,\s)$ exhibits a number of remarkable features:

\begin{itemize}
\item
$U_{\rm eff}(\vp,\s)$ possesses two (infinitely) large flat regions
as a function of $\vp$ at $\s_a = {\rm fixed}$.
\item
The first one -- the (-) flat ``inflaton'' region for large negative values
of $\vp$ (and $\s_a$ -- finite) corresponds to the ``slow-roll'' inflationary 
evolution of the ``early'' universe driven by $\vp$ where:
\be
U_{\rm eff}\bigl(\phi,\s\bigr) \simeq U_{(-)} = 
\frac{M_1^2}{4\chi_2\,M_2} + 2\L_0 \; ,
\lab{U-minus}
\ee
independent of the finite value of $\s_a$, which is energy scale of the 
inflationary epoch. 
Thus, in the ``early'' universe
there is {\em no spontaneous breaking} of electroweak $SU(2)\times U(1)$ symmetry.
Moreover, $\s_a$ does not participate in the ``slow-roll'' inflationary evolution, 
so $\s$ stays constant there equal to the ``false''vacuum value $\s=0$
\ct{Benisty:2020vvm}.
\item
The second flat region is the (+) flat ``inflaton'' region for large positive 
values of $\vp$ (and $\s_a$ -- finite) which corresponds to the evolution of the 
post-inflationary (``late'') universe. Here:
\be
U_{\rm eff}\bigl(\vp,\s\bigr) \simeq U_{(+)}(\s)
=\frac{\Bigl( m^2_0\, \s^{*}_a \s_a - f_1\Bigr)^2}{4\chi_2 f_2} + 2\L_0 
\lab{higgs-potential}
\ee
\begin{figure}[t!]
 	\centering
\includegraphics[width=0.75\textwidth]{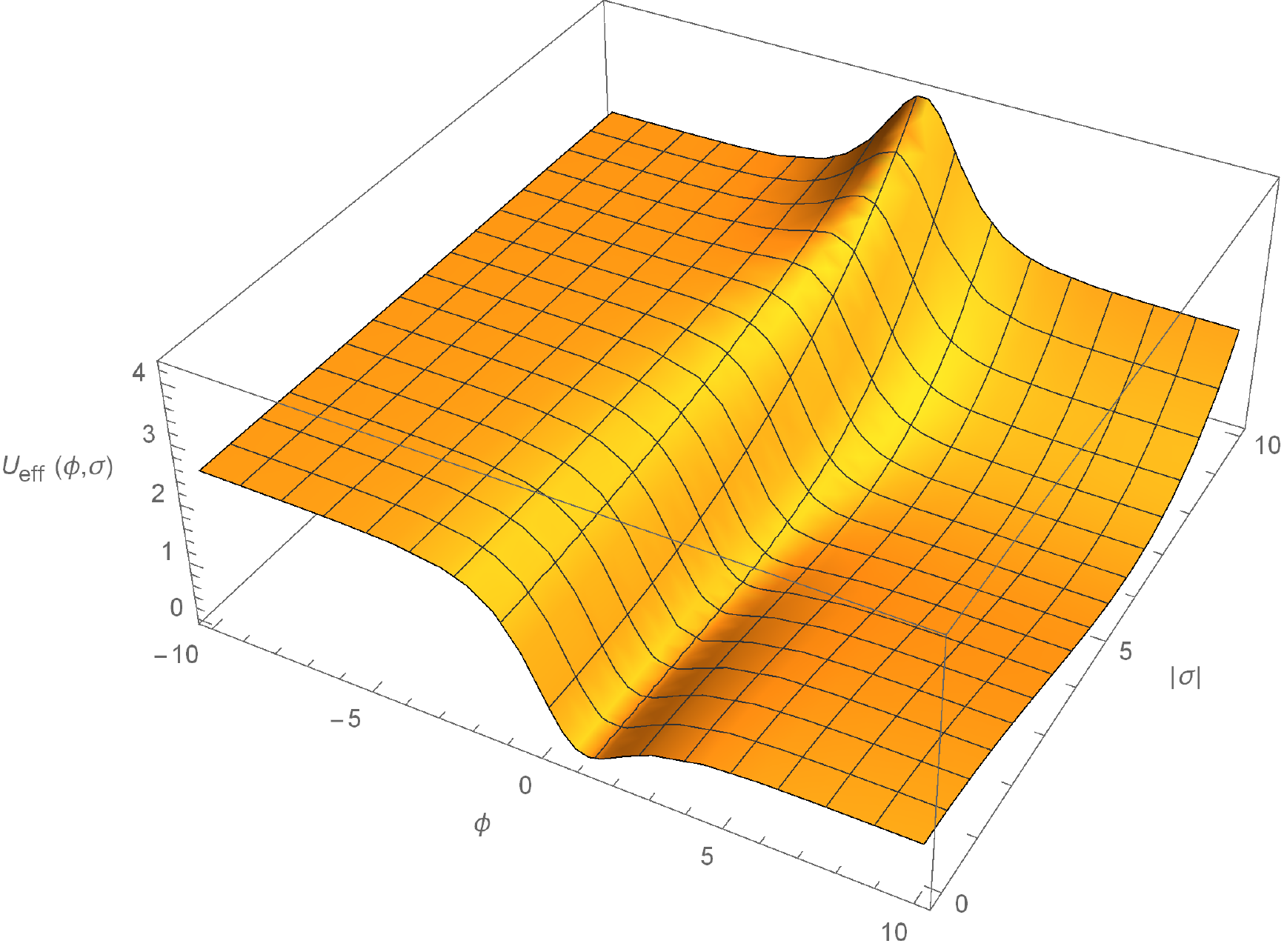}
\caption{\it{Qualitative shape of the two-dimensional plot for the effective scalar potential $U_{eff}$.}}
\end{figure}

becomes a {\em dynamically induced} $SU(2)\times U(1)$ spontaneous
symmetry breaking Higgs-like potential with a Higgs ``vacuum'' at
$|\s_{\rm vac}|=\frac{1}{m_0}\sqrt{f_1}$.
\item
Relations \rf{U-minus}-\rf{higgs-potential} allow the following natural
identification of the scales of the parameters: $\L_0 \sim 10^{-122}
M^4_{Pl}$ (current epoch observable cosmological constant); 
$f_1 \sim f_2 \sim M^4_{EW}$ and $m_0 \sim M_{EW}$ ($M_{EW}$ being the
electroweak mass scale); $M_1 \sim M_2 \sim 10^{-8} M_{Pl}^4$ corresponding to the
``early'' universe's energy scale of inflation being of order $10^{-2} M_{Pl}$.  
\end{itemize}

Concerning confrontation with the observational data, the viability of the 
present model (in a slightly simplified form without the Higgs scalar, 
which as already mentioned does not influence the slow-roll inflationary dynamics) 
has been analyzed and confirmed numerically in Ref.\ct{Guendelman:2014bva}. 
The results for the tensor-to-scalar ratio $r \simeq 10^{-3}$ 
and for the scalar spectral index $n_s \simeq 0.96$ which are in a good 
agreement with the latest 
Further detailed numerical studies on the NRVF models have been presented in
Refs. \cite{Staicova:2016pfd,Staicova:2018bdy,Staicova:2019ksr}.

Let us also note that Ref.\cite{Guendelman:2014bva}
(for an earlier version, see \ct{Guendelman:2002js}) exhibits an explicit
realization of the cosmological ``seesaw'' mechnaism through the NRVF
formulation, as well as it yields an additional ``emergent universe''
cosmological solution without a ``Big-Bang'' initial singularity. For a
brief illustration of the latter effects let us consider the ``inflaton-only''
NRVF action studied in \ct{Guendelman:2014bva} (for simplicity we skip the $R^2$ term):
\br
S = \int d^4 x\, \P_1(A) \Bigl\lb R + X_\vp - f_1 e^{-\a\vp}\Bigr\rb
+ \int d^4 x\, \P_2(B) \Bigl\lb -b\,e^{-\a\vp}\,X_\vp + f_2 e^{-2\a\vp}
- \frac{\P_0 (C)}{\sqrt{-g}}\Bigr\rb \; ,
\lab{NRVF-emergent}
\er
where $b$ is an additional dimensionless parameter.

The ``inflaton'' potential in the Einstein frame (analog of \rf{U-eff}) 
is:
\begin{equation}
U_{\rm eff}(\vp)= \frac{1}{4\chi_2}\bigl(f_1 e^{-\a\vp} + M_1\bigr)^2
\bigl(f_2 e^{-2\a\vp} + M_2\bigr)^{-1}    
\end{equation}
so that on the (-) and (+)
``inflaton'' flat regions $U_{\rm eff}(\vp)$ reduces to:
$U_{(-)} \simeq \frac{f_1^2}{4\chi_2\,f_2}$ and
$U_{(+)} \simeq \frac{M_1^2}{4\chi_2\,M_2}$, accordingly. Therefore, choosing
$f_1 \sim f_2 \sim 10^{-8} M^4_{Pl}$ conforming to the inflationary scale,
and taking $M_1 \sim M^4_{EW}$ and $M_2 \sim  M^4_{Pl}$ we achieve 
$U_{(+)}\sim 10^{-122} M^4_{Pl}$ vastly smaller than $U_{(-)}$. If we take
$\a \to - \a$ in \rf{NRVF-emergent} the roles of $f_{1,2}$ and $M_{1,2}$ are
interchanged.

Similar ``seesaw'' effect is found in Refs.\cite{Benisty:2019vej,Benisty:2020xqm}
where the scalar potential is extracted from the slow-roll parameters.
\footnote{The paper \cite{Benisty:2020xqm} was awarded second prize in the 2020 Essay 
Competition of the Gravity Research Foundation.}. Furthermore, the NRVF model \rf{NRVF-emergent} yields in
EInstein-frame ``emergent universe'' solution for the range of the $b$-parameter: $-4(2-\sqrt{3}) < b \frac{f_1}{f_2} < -1$.
\begin{figure}[t!]
 	\centering
\includegraphics[width=0.75\textwidth]{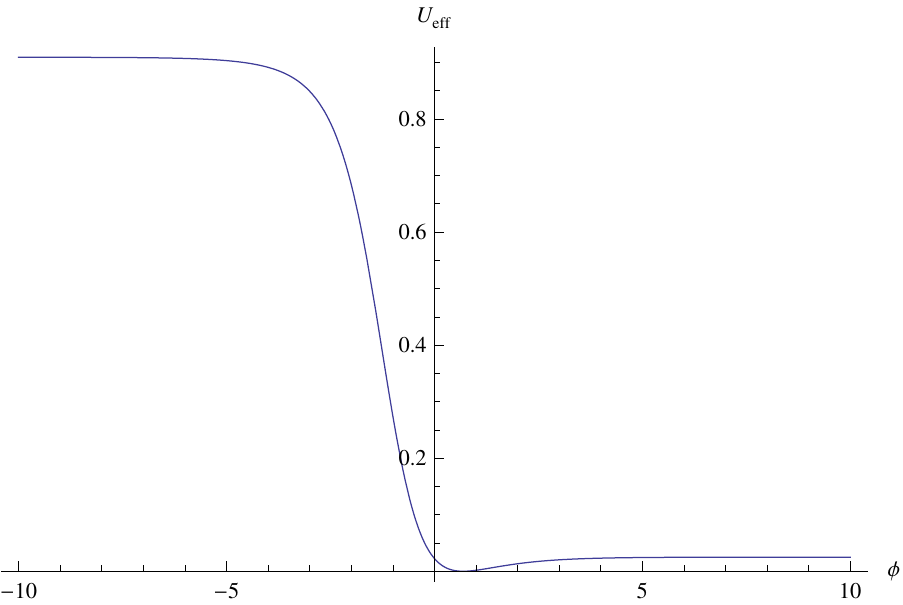}
\caption{\it{Qualitative shape of the one-dimensional plot for the effective scalar potential $U_{eff}$.}}
\end{figure}

\section{Dynamical Generation of Inflation in Metric Formulation}

Let us now consider a substantionally truncated version of the model
\rf{NRVF-2} without any matter fields, 
involving few non-Riemannian volume elements \ct{Benisty:2019tno}:
\be
S = \int d^4 x \Bigl\{\P_1 (A)\Bigl\lb R(g) 
- 2\L_0 \frac{\P_1 (A)}{\sqrt{-g}}\Bigr\rb 
+ \P_2 (B) \frac{\P_0 (C)}{\sqrt{-g}}\Bigr\} \; , 
\lab{NRVF-3}
\ee
where now unlike \rf{NRVF-2} $R(g) \equiv g^{\m\n} R_{\m\n}\bigl(\G(g)\bigr)$ is 
the scalar curvature in the second order (metric) formalism ($\G^\l_{\m\n}(g)$
being the Levi-Civita connection w.r.t. $g_{\m\n}$).

The equations of motion w.r.t. auxiliary tensor gauge fields $A_{\m\n\l}$, $\P_2 (B)$ and 
$\P_1 (C)$ are special cases of the dynamical constraint Eqs.\rf{M1}-\rf{chi2}
with all matter field terms being zero, which again introduce the three free
integration constants $M_{1,2},\,\chi_2$.

Passage to the physical Einstein frame is again realized via the conformal transformation \rf{g-bar}, however this time we have to use the well-known formulas for conformal transformations within the metric formalism (\ct{Dabrowski:2008kx}; bars indicate magnitudes in the ${\bar g}_{\m\n}$-frame):
\br
R_{\m\n}(g) = R_{\m\n}(\bar{g}) - 3 \frac{{\bar g}_{\m\n}}{\chi_1}
{\bar g}^{\k\l} \pa_\k \chi_1^{1/2} \pa_\l \chi_1^{1/2} 
+ \chi_1^{-1/2}\bigl({\bar \nabla}_\m {\bar \nabla}_\n \chi_1^{1/2} +
{\bar g}_{\m\n} {\bar{\Box}}\chi_1^{1/2}\bigr)\; ,\lab{dabrowski-1} 
\er
with $\chi_1 \equiv \frac{\P_1 (A)}{\sqrt{-g}}$. Redefining $\chi_1$ as 
$\chi_1 = \exp{\bigl(u/\sqrt{3}\bigr)}$ allows to write the Einstein-frame NRVF action in
the form:
\br
S_{\rm EF} = \int d^4 x \sqrt{-{\bar g}} \Bigl\lb R({\bar g}) 
- \h {\bar g}^{\m\n}\pa_\m u \pa_\n u - U_{\rm eff} (u) \Bigr\rb\; ,
\lab{EF-action-2} \\
U_{\rm eff} (u) = 2 \L_0 - M_1 \exp{\bigl(-\frac{u}{\sqrt{3}}\bigr)} 
+ \chi_2 M_2 \exp{\bigl(-2 \frac{u}{\sqrt{3}}\bigr)} \; .
\lab{U-eff-2}
\er

Thus, from the original pure-gravity NRVF action \rf{NRVF-3} we derived a
physical Einstein-frame action \rf{EF-action-2}-\rf{U-eff-2} containing
a {\em dynamically created} scalar field $u$ with a non-trivial effective scalar 
potential $U_{\rm eff}(u)$ \rf{U-eff-2} entirely {\em dynamically generated} 
by the initial non-Riemannian volume elements in \rf{NRVF-3} because of the
appearance of the free integration constants $M_{1,2},\,\chi_2$ in their
respective equations of motion. There are two main features of the  effective potential \rf{U-eff-2} which are relevant for cosmological applications with the dynamically created field $u$ as an ``inflaton''.

\begin{itemize}
\item
$U_{\rm eff} (u)$ \rf{U-eff-2} possesses one flat region for large positive values 
of $u$ where $U_{\rm eff} (u) \simeq 2\L_0$, which corresponds to 
``early'' universe' inflationary evolution with energy scale $2\L_0$.
\item
$U_{\rm eff} (u)$ \rf{U-eff-2} has a stable minimum for a small finite value $u=u_{*}$ where $e^{-u_{*}/\sqrt{3}} = M_1/(2\chi_2 M_2)$.
\item
The region around the stable minimum at $u=u_{*}$ correspond to ``late'' universe' evolution where the minimum value of the potential:
\be
U_{\rm eff} (u_{*})= 2\L_0 - \frac{M_1^2}{4\chi_2 M_2} \equiv 2 \L_{\rm DE}
\lab{DE-value}
\ee
is the dark energy density value. 
\end{itemize}

\textbf{Remark.} The effective potential $U_{\rm eff}$
\rf{U-eff-2} generalizes the well-known Starobinsky
inflationary potential \ct{Starobinsky:1979ty} (\rf{U-eff} reduces to Starobinsky
potential upon taking the following special values for the
parameters: $\L_0 = \frac{1}{4}M_1 = \h \chi_2 M_2$). 

In Ref.\ct{Benisty:2019tno} a thorough analysis has been performed of the slow-roll
inflationary dynamics driven by the dynamically created ``inflaton'' $u$
with its dynamically generated effective potential \rf{U-eff-2}, including
explicit calculation of the standard slow-roll parameters $\epsilon$ and
$\eta$, as well we have obtained explicit expressions for the
tensor-to-scalar ratio $r$ and the scalar spectral index $n_s$ of density
perturbations as functions of the number of e-folds $\cN = \log a$ ($a$
being the Friedmann scale factor):
\br
r \simeq \frac{12}{\Bigl\lb \cN + \frac{\sqrt{3}}{4} u_i(\cN) + c_0\Bigr\rb^2}
\lab{ns-r-approx},\quad
n_s \simeq 1 -\frac{r}{4}-\sqrt{\frac{r}{3}},
\er
with $c_0 \equiv \frac{\sqrt{3}}{2} - \frac{3}{4} \log\Bigl(2\bigl(1+2/\sqrt{3}\bigr)\Bigr)$. $u_i (\cN)$ is the value of the ``unflaton'' at the start of inflation
as function of $\cN$.

For a plausible assumption about the scales of $M_{1,2},\,\chi_2$ and taking 
$\cN = 60$ $e$-folds till end of inflation the observables are predicted to be:
$ n_s \approx 0.969\;\;, \;\; r \approx 0.0026$, 
which conform to the \textsl{PLANCK} constraints \ct{Akrami:2018odb}
($0.95 < n_s < 0.97\;\;, \;\; r < 0.064$).
\section{ Dynamical Spacetime Formulation}

Let us now observe that the non-Riemannian volume element $\Omega=\Phi(B)$
\rf{Phi-D} on a Riemannian manifold can be rewritten using Hodge duality
(here $D=4$) in terms of a vector field
$\chi^\m = \frac{1}{3!} \frac{1}{\sqrt{-g}} \vareps^{\m\n\k\l} B_{\n\k\l}$
so that $\Omega$ becomes $\Omega (\chi) = \pa_\m \bigl(\sqrt{-g}\chi^\m\bigr)$,
~\textsl{i.e.} it is a non-canonical volume element different from $\sqrt{-g}$, 
but still involving the metric. It can be represented alternatively through a
Lagrangian multiplier action term yielding covariant conservation of a
specific energy-momentum tensor of the form $\cT^{\m\n} = g^{\m\n} \cL$:
\be
\mathcal{S}_{(\chi)}=\int d^{4}x\sqrt{-g} \, \chi_{\mu;\nu}\cT^{\mu\nu}
= \int d^4 x \pa_\m \bigl(\sqrt{-g}\chi^\m\bigr) \bigl(-\cL\bigr) \; ,
\lab{action}
\ee
where $\chi_{\m;\n}=\pa_\n\chi_\m -\G_{\m\n}^\l\chi_\l$. The vector field $\chi_\mu$ is called {\em ``dynamical space time vector''}, because of the energy density of $\cT^{00}$ is a canonically conjugated momentum 
w.r.t. $\chi_0$, which is what we expected from a dynamical time. 

In what follows we will briefly consider a new class of gravity-matter 
theories based on the ordinary Riemannian volume element $\sqrt{-g}$ but
involving action terms of the form \rf{action} where now $\cT^{\m\n}$ is of
more general form than $\cT^{\m\n} = g^{\m\n} \cL$. This new formalism is called {\em ``dynamical spacetime formalism''} \ct{Guendelman:2009ck,Benisty:2016ybt} due to the above remark on $\chi_0$. 

Ref.\cite{Benisty:2018qed} describes a unification between dark energy and dark 
matter by introducing a quintessential scalar field in addition to the dynamical 
time action. The total Lagrangian reads:
\be
\mathcal{L}=\frac{1}{2}R+\chi_{\mu;\nu}\cT^{\mu\nu} - \frac{1}{2}g^{\alpha\beta} \phi_{,\alpha}\phi_{,\beta} - V(\phi),
\lab{action-with-xi}
\ee
with energy-momentum tensor 
$\cT^{\mu\nu} = -\frac{1}{2} \phi^{,\mu} \phi^{,\nu}$. 
From the variation of the Lagrangian term $\chi_{\mu;\nu}\cT^{\mu\nu}$ with respect to the
vector field $\chi_{\mu}$, the covariant conservation of the energy-momentum tensor 
$\nabla_{\mu} \cT^{\m\n} = 0$ is implemented. The latter within the FLRW
framework forces the kinetic term of the scalar field to behave
as a dark matter component:
\begin{equation}\label{var1}
\nabla_{\mu}\cT^{\mu\nu} = 0 \quad \Rightarrow \quad \dot{\phi}^2 = \frac{2 \Omega_{m0}}{a^3}.
\end{equation}
where $\Omega_{m0}$ is an integration constant. 
The variation with respect to the scalar field $\phi$ yields a current: 
\begin{equation}\label{var22}
- V'(\phi) = \nabla_\mu j^{\mu} , \quad j^{\mu}  = \frac{1}{2}\phi_{,\nu} (\chi^{\mu;\nu}+\chi^{\nu;\mu}) + \phi^{,\mu}
\end{equation}
For constant potential $V(\phi) = \Omega_\Lambda = const$ the current is covariantly conserved. In the FLRW setting, where the dynamical time ansatz introduces only a time component $\chi_\mu =(\chi_0,0,0,0)$, 
the variation (\ref{var22}) gives:
\begin{equation}\label{var2}
 \dot{\chi}_0 - 1 = \xi \,  a^{-3/2}  ,
\end{equation}
where $\xi$ is an integration constant. Accordingly, the FLRW energy density and
pressure read:
\begin{equation}
\label{setvv}
\rho = \left(\dot{\chi}_0-\frac{1}{2}\right)\dot{\phi}^2 + V, \quad p = \frac{1}{2}\dot{\phi}^2(\dot{\chi}_0-1) - V.
\end{equation}
Plugging the relations (\ref{var1},\ref{var2}) into the density and the pressure 
terms (\ref{setvv}) yields the following simple form of the latter:
\be
\rho = \Omega_\Lambda +\frac{\xi \Omega_{m 0}}{a^{9/2}} +\frac{\Omega_{m 0}}{a^3}, \quad p = -\Omega_\Lambda+ \frac{\xi \Omega_{m0}}{2\, a^{9/2}}.
\lab{Frid}
\ee
In \rf{Frid} there are 3 components for the "dark fluid": dark energy with 
$\omega_\Lambda = -1$, dark matter with $\omega_m = 0$ and an additional equation 
of state $\omega_\xi = 1/2$.
For non-vanishing and negative $\xi$ the additional part introduces a minimal 
scale parameter, which avoids singularities. If the dynamical time is equivalent 
to the cosmic time $\chi_0 = t$, we obtain $\xi = 0$ from Eq.(\ref{var2}), 
whereupon the density and the pressure terms \rf{Frid} coincide with those from the $\Lambda$CDM model precisely. The additional part (for $\xi \neq 0$) fits to the late time accelerated expansion data \cite{Anagnostopoulos:2019myt}.
\begin{figure}[t!]
 	\centering
\includegraphics[width=0.8\textwidth]{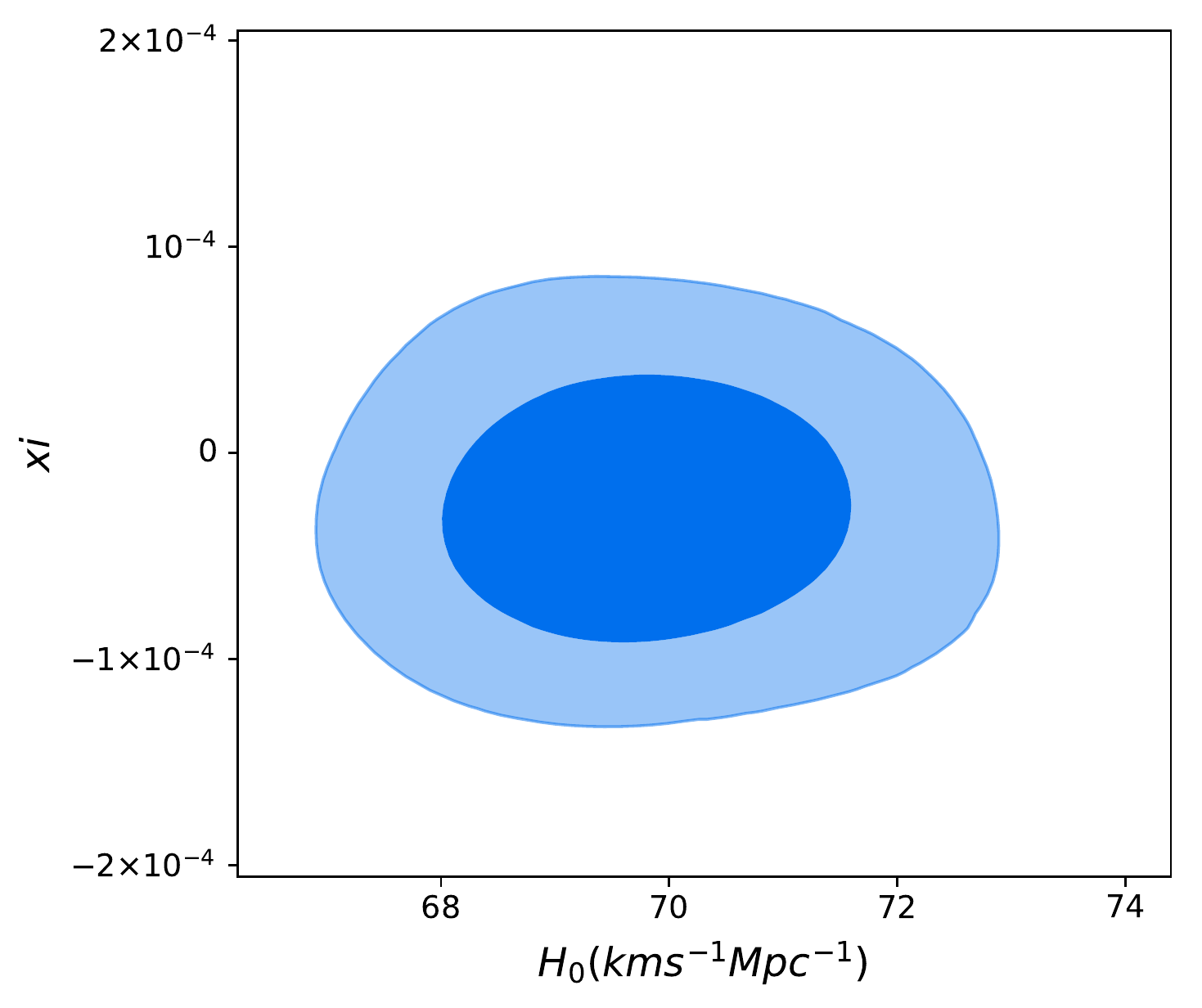}
\caption{\it{The posterior distribution for the simplest case of DST cosmology in the parameter space $\xi$ vs. $H_0$.}}
\end{figure}
 {In order to constraint our model, we deploy the following data sets: \textbf{Cosmic Chronometers (CC)} exploit the evolution of differential ages of passive galaxies at different redshifts to directly constrain the Hubble parameter \cite{Jimenez:2001gg}. We use uncorrelated 30 CC measurements of $H(z)$ discussed in \cite{Moresco:2012by,Moresco:2012jh,Moresco:2015cya,Moresco:2016mzx}. As \textbf{Standard Candles (SC)} we use uncorrelated measurements of the Pantheon Type Ia supernova \cite{Scolnic:2017caz} that were collected in \cite{Anagnostopoulos:2020ctz}. In addition, we use the uncorrelated data points from different \textbf{Baryon Acoustic Oscillations (BAO)} collected in \cite{Benisty:2020otr} from \cite{Percival:2009xn,Beutler:2011hx,Busca:2012bu,Anderson:2012sa,Seo:2012xy,Ross:2014qpa,Tojeiro:2014eea,Bautista:2017wwp,deCarvalho:2017xye,Ata:2017dya,Abbott:2017wcz,Molavi:2019mlh}. Studies of the BAO features in the transverse direction provide a measurement of $D_H(z)/r_d = c/H(z)r_d$, with the comoving angular diameter distance defined in \cite{Hogg:2020ktc,Martinelli:2020hud}. In our database we use the parameters $D_A = D_M/(1+z)$ and }
\begin{equation}
    D_V(z) \equiv [ z D_H(z) D_M^2(z) ]^{1/3}.
\end{equation}
 {which is a combination of the BAO peak coordinates. $r_d$ is the sound horizon at the drag epoch. Finally, for very precise  "line-of-sight" (or "radial") observations, BAO can also measure directly the Hubble parameter \cite{Benitez:2008fs}.  }

 {The posterior distribution yields: $\Omega_m = 0.2811 \pm 0.0743$, $\Omega_\Lambda = 0.7187 \pm 0.07351$, $\xi = (-2.776 \pm 3.023)\cdot 10^{5}$ and with the Hubble parameter $H_0 = 69.72 \pm 1.239 Mpc/(km/sec)$. \cite{Benisty:2018gzx} shows that with higher dimensions, the solution derived from the Lagrangian \rf{action-with-xi} describes inflation, where the total volume oscillates and the original scale parameter exponentially {\em grows}. }

The dynamical spacetime Lagrangian can be generalized to yield  a {\em diffusive energy-momentum tensor}. Ref. \cite{Calogero:2013zba} shows that
the diffusion equation has the form:
\begin{equation}\label{diffusion}
\nabla_\mu \cT^{\mu\nu}=3\sigma j^\nu, \quad j^{\mu}_{;\mu}=0, \end{equation}
where $\sigma$ is the diffusion coefficient and $j^{\mu}$ is a current source. 
The covariant conservation of the current source indicates the conservation of 
the number of the particles. 
By introducing the vector field $\chi_\mu$ in a different part of the 
Lagrangian:
\begin{equation} \label{nhd1}
\mathcal{L}_{(\chi,A)}=\chi_{\mu;\nu}\cT^{\mu\nu} +\frac{\sigma}{2} (\chi_{\mu}+\partial_{\mu}A)^2, 
\end{equation} 
the energy-momentum tensor $\cT^{\mu\nu}$ {\em gets a diffusive source}.
From a variation with respect to the dynamical 
space time vector field $\chi_{\mu}$ we obtain:
\begin{equation} \label{nhd2}
\nabla_{\nu}\cT^{\mu\nu}=\sigma(\chi^{\mu}+\partial^{\mu}A)= f^\mu,
\end{equation} 
a current source $f^\mu=\sigma (\chi^{\mu}+\partial^{\mu} A )$ for the 
energy-momentum tensor. From the variation with respect to the new scalar $A$,
a covariant conservation of the current emerges $f^\mu_{;\mu}=0$. 
{\em The latter relations correspond to the diffusion equation (\ref{diffusion})}. Refs.\cite{Benisty:2017eqh,Benisty:2017rbw,Benisty:2017lmt,Bahamonde:2018uao} study the cosmological solution using the energy-momentum tensor
$\cT^{\mu\nu} = -\frac{1}{2}g^{\mu\nu} \phi^{,\lambda}\phi_{\lambda}$. 
The total Lagrangian reads:
\begin{equation}
\mathcal{L}=\frac{1}{2}R - \frac{1}{2}g^{\alpha\beta} \phi_{,\alpha}\phi_{,\beta} - V(\phi) \\+\chi_{\mu;\nu}\cT^{\mu\nu} +\frac{\sigma}{2} (\chi_{\mu}+\partial_{\mu}A)^2.
\end{equation}
The FLRW solution unifies the dark energy and the dark matter originating from 
one scalar field with possible diffusion interaction. 
Ref.\cite{Benisty:2018oyy} investigates more general energy-momentum tensor 
combinations and shows that asymptotically all of the combinations yield 
$\Lambda$CDM model as a stable fixed point. \cite{Banerjee:2019kgu} shows that the DST theories and Diffusive extensions can describe a Lagrangian formulation for Running Vacuum Models.

\section{Scale Invariance, Fifth Force in Fermionic and Dust Matter}
 {In this class of theories the fifth force problem can be solved,  this can be checked most simply in a theory with two volume elements (integration measure densities) \ct{Guendelman:1999qt,Guendelman:1999tb}, where at least one of them was a non-canonical one and short-termed ``two-measure theory'' (TMT). The result is expected to be generic however. This model has a number of remarkable properties if fermions are included in a self-consistent way \ct{Guendelman:1999tb}. In this case, the constraint that arises in the TMT models in the Palatini formalism can be represented as an equation for $\chi\equiv\Phi/\sqrt{-g}$, in which the left side has an order of the vacuum energy density, and the right side (in the case of non-relativistic fermions) is proportional 
to the fermion density. Moreover, it turns out that even cold fermions have a (non-canonical) pressure $P_f^{noncan}$ and the corresponding contribution to the energy-momentum tensor has the structure of a  cosmological constant term which is proportional to the fermion density. The  remarkable fact is that the right hand side of the 
constraint coincide with $P_f^{noncan}$. This allows us to construct a cosmological model \cite{Guendelman:2012vc} of the late universe in which dark energy is generated by a gas of non-relativistic neutrinos without the need to introduce into the model a specially designed scalar field.} 

 {In models with a scalar field, the requirement of scale invariance of the 
initial action\cite{Guendelman:1999qt} plays a very constructive role. It allows to construct a model\cite{Guendelman:2006af} where without 
fine tuning we have realized: absence of initial singularity of the curvature; k-essence;  inflation with graceful exit to zero cosmological constant. Of particular interest are scale invariant  models in which both fermions and a dilaton scalar field $\phi$ are present. Then it turns out that the Yukawa coupling of fermions to $\phi$ is proportional to  $P_f^{noncan}$. As a result, 
it follows from the constraint, that in all cases when fermions are in states which constitute a regular barionic matter, the Yukawa coupling of fermions to dilaton has an order of ratio of the vacuum energy density to the fermion energy density \cite{Guendelman:2006ji}. Thus, the theory provides a solution of the 5-th force problem without any fine tuning or a special design of the model. Besides, in the described states, the regular Enstein's equations 
are reproduced. In the opposite case, when fermions are very deluted, e.g. in the model of the late Universe filled with a cold neutrino gas, the neutrino dark energy appears 
in such a way that the dilaton $\phi$ dynamics is closely correlated with that of the neutrino gas \cite{Guendelman:2006ji}.  }

 {Scale invariant model containing a dilaton $\phi$ and dust (as a model of matter)\cite{Guendelman:2007ph} possesses similar features. Dilaton to matter coupling "constant" $f$ appears to be dependent of the matter density. To see this more explicitly, let us consider the action containing a non Riemannian measure $\Phi$ which is a total divergence and is invariant under the global scale transformations:}
\begin{equation}
    g_{\mu\nu}\rightarrow e^{\theta }g_{\mu\nu}, \quad
\Gamma^{\mu}_{\alpha\beta}\rightarrow \Gamma^{\mu}_{\alpha\beta},
\quad \phi\rightarrow \phi-\frac{M_{p}}{\alpha}\theta, \quad
\Phi \rightarrow e^{2\theta}\Phi \label{st}
\end{equation}
 {where $\theta =const$. It is convenient to represent the action in the following form:}
\begin{eqnarray}
S&=&S_g+S_{\phi}+S_{m} \label{SgSphiSm}
\label{totaction}\\
 S_g&=&-\frac{1}{\kappa}\int (\Phi +b_{g}\sqrt{-g})R(\Gamma,g) e^{\alpha\phi
 /M_{p}}d^{4}x \,;
\nonumber\\
S_{\phi}&=&\int e^{\alpha\phi/M_{p}}\left[(\Phi
+b_{\phi}\sqrt{-g})\frac{1}{2}g^{\mu\nu}\phi_{,\mu}\phi_{,\nu}-
 \left(\Phi V_{1}
+\sqrt{-g}V_{2}\right)e^{\alpha\phi /M_{p}}\right]d^{4}x \,;
\nonumber\\
S_{m}&=&\int (\Phi +b_{m}\sqrt{-g})L_m d^{4}x \, , \nonumber
\nonumber
\end{eqnarray}
 {where the Lagrangian for the matter, as collection of particles, which provides the scale invariance of $S_m$ reads}
\begin{equation}
L_m=-m\sum_{i}\int e^{\frac{1}{2}\alpha\phi/M_{p}}
\sqrt{g_{\alpha\beta}\frac{dx_i^{\alpha}}{d\lambda}\frac{dx_i^{\beta}}{d\lambda}}\,
\frac{\delta^4(x-x_i(\lambda))}{\sqrt{-g}}d\lambda \label{Lm}
\end{equation}
 {where $\lambda$ is an arbitrary parameter. For simplicity we consider the collection of the particles with the same mass parameter $m$. We assume in addition that $x_i(\lambda)$ do not participate in the scale transformations (\ref{st}). We will assume that
$d\vec{x}_i/d\lambda \equiv 0$  for all particles. It is
convenient to proceed in the frame where $g_{0l}=0$, \, $l=1,2,3$. Then the particle density is defined by}
\begin{equation}
n(\vec{x})=\sum_{i}\frac{1}{\sqrt{-g_{(3)}}}\delta^{(3)}(\vec{x}-\vec{x}_i(\lambda))
\label{n(x)}
\end{equation}
where $g_{(3)}=\det(g_{kl})$ and
\begin{equation}
S_{m}=-m\int d^{4}x(\Phi
+b_{m}\sqrt{-g})\,n(\vec{x})\,e^{\frac{1}{2}\alpha\phi/M_{p}}
\label{S-n(x)}
\end{equation}
 {It turns out that when working with the new metric ($\phi$
 remains the same)
\begin{equation}
\tilde{g}_{\mu\nu}=e^{\alpha\phi/M_{p}}(\zeta +b_{g})g_{\mu\nu},
\label{ct}
\end{equation}
which we call the Einstein frame,
 the connection  becomes Riemannian.
  Notice that 
$\tilde{g}_{\mu\nu}$ is invariant under the scale transformations (\ref{st}).  The transformation (\ref{ct}) causes the transformation of the particle density}
\begin{equation}
\tilde{n}(\vec{x})=(\zeta +b_g)^{-3/2}\,
e^{-\frac{3}{2}\alpha\phi/M_{p}}\, n(\vec{x}) \label{ntilde}
\end{equation}
 {After the change of variables  to the Einstein frame (\ref{ct}) and some simple algebra, the gravitational equations take the standard GR form}
\begin{equation}
G_{\mu\nu}(\tilde{g}_{\alpha\beta})=\frac{\kappa}{2}T_{\mu\nu}^{eff}
 \label{gef}
\end{equation}
 {where $G_{\mu\nu}(\tilde{g}_{\alpha\beta})$ is the Einstein tensor in the Riemannian space-time with the metric $\tilde{g}_{\mu\nu}$. The components of the effective energy-momentum tensor are as follows:}
\begin{eqnarray}
T_{00}^{eff}&=&\frac{\zeta +b_{\phi}}{\zeta
+b_{g}} \left(\dot{\phi}^2- \tilde{g}_{00}X\right) \label{T00}\\
&+&\tilde{g}_{00}\left[V_{eff}(\phi;\zeta,M)-\frac{\delta\cdot
b_g}{\zeta +b_{g}}X+\frac{3\zeta +b_m +2b_g}{2\sqrt{\zeta
+b_{g}}}\, m\, \tilde{n}\right] \nonumber
\end{eqnarray}
\begin{eqnarray}
T_{ij}^{eff}&=&\frac{\zeta +b_{\phi}}{\zeta +b_{g}}
\left(\phi_{,k}\phi_{,l}-\tilde{g}_{kl}X\right)
\label{Tkl}\\
&+&\tilde{g}_{kl}\left[V_{eff}(\phi;\zeta,M)-\frac{\delta\cdot
b_g}{\zeta +b_{g}}X+\frac{\zeta -b_m +2b_g}{2\sqrt{\zeta
+b_{g}}}\, m\, \tilde{n}\right]\nonumber
\end{eqnarray}
 {Here the following notations have been used:}
\begin{equation}
X\equiv\frac{1}{2}\tilde{g}^{\alpha\beta}\phi_{,\alpha}\phi_{,\beta}
\qquad and \qquad \delta =\frac{b_{g}-b_{\phi}}{b_{g}}
\label{delta}
\end{equation}
and the function $V_{eff}(\phi ;\zeta)$ is defined by
\begin{equation}
V_{eff}(\phi ;\zeta)=
\frac{b_{g}\left[M^{4}e^{-2\alpha\phi/M_{p}}+V_{1}\right]
-V_{2}}{(\zeta +b_{g})^{2}} \label{Veff1}
\end{equation}
 {The dilaton $\phi$ field equation in the Einstein frame is as follows}
\begin{eqnarray}
&&\frac{1}{\sqrt{-\tilde{g}}}\partial_{\mu}\left[\frac{\zeta
+b_{\phi}}{\zeta
+b_{g}}\sqrt{-\tilde{g}}\tilde{g}^{\mu\nu}\partial_{\nu}\phi\right]-\frac{\alpha}{M_{p}}\,\frac{(\zeta
+b_{g})M^{4}e^{-2\alpha\phi/M_{p}}-(\zeta -b_{g})V_{1}
-2V_{2}-\delta b_{g}(\zeta
+b_{g})X}{(\zeta +b_{g})^{2}}\nonumber\\
 &&=\frac{\alpha}{M_{p}}\,\frac{\zeta
-b_m +2b_g}{2\sqrt{\zeta +b_{g}}}\, m\,\tilde{n}
 \label{phief}
\end{eqnarray}
In the above equations, the scalar field $\zeta$  is determined as
a function $\zeta(\phi,X,\tilde{n})$ by means of the following
constraint:
\begin{eqnarray}
&&\frac{(b_{g}-\zeta)\left(M^{4}e^{-2\alpha\phi/M_{p}}+V_{1}\right)-2V_{2}}{(\zeta
+b_g)^2} -\frac{\delta\cdot b_{g}X}{\zeta +b_g}=
\frac{\zeta -b_m +2b_g}{2\sqrt{\zeta +b_{g}}}\, m\, \tilde{n}\nonumber\\
\label{constraint2}
\end{eqnarray}
One should now pay attention to the interesting result that  the
explicit $\tilde{n}$ dependence involving {\bf the  same form of
$\zeta$ dependence}
\begin{equation}
 \frac{\zeta -b_m +2b_g}{2\sqrt{\zeta +b_{g}}}\, m\,
\tilde{n} \label{universality}
\end{equation}
 {appears simultaneously\footnote{Note that analogous result has been observed earlier in the model \cite{Guendelman:2012vc,Guendelman:2006af} where fermionic matter has been studied instead of the macroscopic (dust) matter in the present model.} in the dust contribution to the pressure (through the last term in Eq. (\ref{Tkl})), in the  effective dilaton to dust coupling (in the r.h.s. of Eq. (\ref{phief})) and in the r.h.s. of the constraint (\ref{constraint2}).}

 {Let us analyze consequences of this wonderful coincidence in the case when the matter energy density (modeled by dust) is much larger than the dilaton contribution to the dark energy density in the space region occupied by this matter. Evidently this is the condition under which all tests of Einstein's GR, including the question of the fifth force, are fulfilled. if the dust is in the normal conditions there is a possibility to provide the desirable feature of the dust in GR: it must be pressureless. This is realized provided that in normal  conditions (n.c.) the following equality holds with extremely high accuracy:}
\begin{equation}
 \zeta^{(n.c.)}\approx b_m-2b_g
\label{decoupling-cond}
\end{equation}
Remind that we have assumed  $b_m >b_g$. Then $\zeta^{(n.c.)}+b_g
>0$, and the transformation (\ref{ct}) and the subsequent
equations in the Einstein frame are well defined.
 Inserting
(\ref{decoupling-cond}) in the last term of Eq. (\ref{T00}) we
obtain the effective dust energy density in normal conditions
\begin{equation}
 \rho_m^{(n.c.)}=2\sqrt{b_m-b_g} \, m\tilde{n}
\label{rho-m-n.c.}
\end{equation}
 {When we get only a slight deviation of from $\zeta$ from $b_m-2b_g$, when the matter energy density is many orders of magnitude larger than the dilaton contribution to the dark energy density, we obtain an effective 5th force  coupling  $f$. For this look at the $\phi$-equation in the form (\ref{phief}) and estimate the Yukawa type coupling constant in the r.h.s. of this equation. In fact, using the constraint (\ref{constraint2}) and representing the particle density in the form $\tilde{n}\approx N/\upsilon$ where $N$ is the number of
particles in a volume $\upsilon$, one can make the following estimation for the effective dilaton to matter coupling "constant" $f$ defined by the Yukawa type interaction term $f\tilde{n}\phi$ (if we were to invent an effective action whose variation with respect to $\phi$ would result in Eq. (\ref{phief})):}
\begin{equation}
f \equiv\alpha\frac{m}{M_{p}}\,\frac{\zeta -b_m
+2b_g}{2\sqrt{\zeta +b_{g}}}\approx
\alpha\frac{m}{M_{p}}\,\frac{\zeta -b_m
+2b_g}{2\sqrt{b_m-b_{g}}}\sim
\frac{\alpha}{M_{p}}\,\frac{\rho_{vac}}{\tilde{n}} \approx
\alpha\frac{\rho_{vac}\upsilon}{NM_{p}} \label{Archimed}
\end{equation}
 {becomes less than the ratio of the "mass of the vacuum" in the volume occupied by the matter to the Planck mass. The model yields this kind  of "Archimedes law" without any especial (intended for this) choice of the underlying action and without fine tuning of the parameters. The model not only explains why all attempts to discover a scalar force correction to Newtonian gravity were unsuccessful so far but also predicts that in the near future there is no chance to detect such corrections  in the astronomical measurements as well as in the specially designed  fifth force experiments on intermediate, short (like millimeter) and even ultrashort (a few nanometer) ranges. This prediction is alternative to predictions of other known models.}

 {Finally, we want to point out fundamental differences of our solution of the fifth force force problem to the Chameleon approach. The important point to make is that we are talking of totally different mechanisms, in the Chameleon model, the proposed quintessential scalar, the Chameleon field has a mass in vacuum which is very small, of the order of the Hubble parameter for example (or in any case very very small). The Chameleon scalar however becomes massive in presence of dense matter,  in compact objects, like Earth, a typical number for this mass has been cited, 
$ m^{-1}\sim 60_{mm}$ \cite{Khoury:2003aq}. This is why a quanta of this scalar field  can penetrate only into a thin shell of the body in the depth about 60micrometer, and the fifth force acts only on the thin shell. This is a way the Chameleon model is argued to explain the smallness of the fifth force. In our case there is no mass generation whatsoever since for our dilaton field, what happens here is the vanishing of the effective coupling constant between the dilaton field and the dense matter, while the dilaton keeps is mass zero or very close to zero. The elimination of interaction between our dilaton field and dense matter is total and absolute, in comparison, a Chameleon wave can  suffer a total reflection from a dense matter region, in such a situation it will not be a total elimination of the fifth force, but it may be hard indeed to prepare such an experiment. The elimination of the fifth force in the Chameleon model is argued to exist because in a spherically symmetric static configuration of a macroscopic object only a very small shell of the object can be a source of the Chameleon scalar, while in our case there would be no source for the scalar, not even the edge or surface of the dense object or at any place of the dense object. Higher-order theories of gravity also have been also studied in connection of fifth force suppression and have been shown to produce an explicit realization of the Chameleon scenario from first principles \cite{Capozziello:2007eu,Capozziello:2012ie}. }

\section*{Acknowledgements} 
We all are grateful for support by COST Action CA-15117 (CANTATA), COST Action CA-16104 and COST Action CA-18108. D.B. thanks Ben-Gurion University of the Negev and 
Frankfurt Institute for Advanced Studies for generous support. E.N. and S.P. are partially supported by Bulgarian National Science Fund Grant DN 18/1.

\bibliographystyle{unsrt}
\bibliography{ref}

\end{document}